\documentclass[english,twocolumn,aps,superscriptaddress,nofootinbib]{revtex4-1}

\usepackage[T1]{fontenc}
\usepackage[latin9]{inputenc}
\usepackage{color}
\usepackage{amsmath,amsthm,amssymb}
\usepackage{graphicx}
\usepackage{esint}
\usepackage{titlesec}
\newlength{\titleskip}
\titleformat{name=\section,numberless}[block]{\large\sffamily\bfseries}{\thesection}{0em}{}
\titlespacing*{name=\section,numberless}{0em}{9pt}{-\parskip}

\newif\iftwocolumns
\twocolumnstrue

\begin{document}

\title{Microscopic theory and quantum simulation of atomic heat transport }

\author{Aris Marcolongo}
\affiliation{
  SISSA -- Scuola Internazionale Superiore di Studi Avanzati, Via
  Bonomea 265, 34136 Trieste, Italy
}

\author{Paolo Umari}
\affiliation{
  Dipartimento di Fisica e Astronomia, Universit\`a di Padova, Via
  Marzolo 8, I-35131 Padova, Italy
}

\author{Stefano Baroni}
\iftwocolumns\email{Corresponding author: baroni@sissa.it}\fi
\affiliation{
  SISSA -- Scuola Internazionale Superiore di Studi Avanzati, Via
  Bonomea 265, 34136 Trieste, Italy
}

\begin{abstract}
  Quantum simulation methods based on density-functional theory are
  currently deemed unfit to cope with atomic heat transport within the
  Green-Kubo formalism, because quantum-mechanical energy densities
  and currents are inherently ill-defined at the atomic scale. We show
  that, while this difficulty would also affect classical simulations,
  thermal conductivity is indeed insensitive to such ill-definedness
  by virtue of a sort of \emph{gauge invariance} resulting from energy
  extensivity and conservation. Based on these findings, we derive an
  expression for the adiabatic energy flux from density-functional
  theory, which allows heat transport to be simulated using ab-initio
  equilibrium molecular dynamics. Our methodology is demonstrated by
  comparing its predictions with those of classical equilibrium and
  ab-initio non-equilibrium (M\"uller-Plathe) simulations for a
  liquid-Argon model, and finally applied to heavy water at ambient
  conditions.
\end{abstract}

\maketitle

\section*{Introduction} Understanding heat transport is key in many
fields of science and technology, such as materials and planetary
sciences, energy saving, heat dissipation and shielding, or
thermoelectric conversion, to name but a few. Heat transport in
insulators is determined by the dynamics of atoms, the electrons
following adiabatically in their ground state.  Simulating atomic heat
transport usually relies on Boltzmann's kinetic approach
\cite{Klemens:1959}, or on molecular dynamics (MD), both in its
equilibrium (Green-Kubo, GK \cite{Green:1954,Kubo,Evans-Morriss,%
  Allen-Tildesley}) and non-equilibrium \cite{Evans-Morriss,%
  Allen-Tildesley,Mueller-Plathe:1997} flavors. The Boltzmann equation
only applies to crystalline solids well below melting, whereas
classical MD (CMD) bears on those materials and conditions that can be
modeled by inter-atomic potentials. Equilibrium ab-initio (AI) MD
\cite{Car:1985,Marx-Hutter} is set to overcome these limitations, but
it is still surprisingly thought to be unfit to cope with thermal
transport \emph{because in first-principles calculations it is
  impossible to uniquely decompose the total energy into individual
  contributions from each atom} (excerpted from Ref.
\onlinecite{Stackhouse:2010}). Such a unique decomposition is not
possible in classical mechanics either, because the potential energy
of a system of interacting atoms can be partitioned into local
contributions in an infinite number of equivalent ways. The quantum
mechanical energy density is also affected by a similar indeterminacy.
Notwithstanding, the expression for the heat conductivity derived from
any sensible energy partitioning or density should obviously be well
defined, as any measurable quantity must.

In this work we first demonstrate that the thermal conductivity
resulting from the GK relation is unaffected by the indeterminacy of
the microscopic energy density; we then introduce a form of energy
density, and a corresponding adiabatic energy flux, from which heat
transport coefficients can be computed within the GK formalism, using
density-functional theory (DFT). Our approach is validated by
comparing the results of equilibrium AIMD with those of
non-equilibrium (M\"uller-Plathe, MP \cite{Mueller-Plathe:1997}) AIMD
and equilibrium CMD simulations for a liquid-Argon model, for which
accurate inter-atomic potentials are derived by matching the forces
generated by them with quantum-mechanical forces computed along the
AIMD trajectories. The case of molecular fluids is finally addressed,
and illustrated in the case of water at ambient conditions.

\section*{Theory}
According to the GK theory \cite{Green:1954,Kubo}, the atomic thermal
conductivity of an isotropic system is given by:
\begin{equation}
  \kappa = \frac{1}{3Vk_{B}T^{2}}\int_{0}^{\infty}\langle
  \mathbf{J}_{q}(t)\cdot\mathbf{J}_{q}(0) \rangle
  dt, \label{eq:Green-Kubo} 
\end{equation}
where brackets $\langle\cdot\rangle$ indicate canonical averages,
$k_{B}$ is the Boltzmann constant, $V$ and $T$ the system volume and
temperature,
$\mathbf{J}_{q}(t) = \int\bigl(\mathbf{j}_{e}(\mathbf{r},t) +
\left(p+\langle e\rangle \right) \mathbf{v}(\mathbf{r},t) \bigr)
d\mathbf{r}$
is the macroscopic heat flux, $\mathbf{j}_{e}$, $\mathbf{v}$, $p$, and
$\langle e\rangle$ being the energy-current density, local velocity
field, and equilibrium values of pressure and energy density,
respectively \cite{Kadanoff-Martin,Forster}.  For further reference,
we define as \emph{diffusive} a flux that results in a non-vanishing
GK conductivity, according to Eq. \eqref{eq:Green-Kubo}.  The integral
of the velocity field is non diffusive in solids and can be assumed to
vanish in one-component fluids, because of momentum conservation. In
these cases, as well as in molecular fluids as we will see, we can
therefore assume that heat and energy fluxes coincide.

Energy is extensive: it can thus be expressed as the integral of a
density, which is defined up to the divergence of a bounded vector
field: two densities that differ by such a divergence, $e(\mathbf{r})$
and
$e'(\mathbf{r})=e(\mathbf{r})+ \partial\cdot \mathbf{p}(\mathbf{r})$,
are indeed equivalent, in that their integrals differ by a surface
term, which is irrelevant in the thermodinamic limit, and can thus be
thought of as \emph{different gauges} of a same scalar field.  Energy
is also conserved: therefore, for any given gauge of its density, $e$,
a corresponding current density, $\mathbf{j}_{e}$, can be defined so
as to satisfy the continuity equation:
\begin{equation}
  \dot{e}(\mathbf{r},t)
  + \partial\cdot\mathbf{j}_{e}(\mathbf{r},t)=0. \label{eq:continuity} 
\end{equation}
According to Eq. \eqref{eq:continuity} the macroscopic fluxes in two
different energy gauges differ by a total time derivative, which is
non-diffusive:
$\mathbf{J}'_{e}(t)=\mathbf{J}_{e}(t)+\dot{\mathbf{P}}(t)$, where
$\mathbf{P}(t)=\int\mathbf{p}(\mathbf{r},t)d\mathbf{r}$.  The equality
of the corresponding heat conductivities results from the following

\emph{Lemma}---Let $\mathbf{J}_{1}$ and $\mathbf{J}_{2}$ be two
macroscopic fluxes defined for a same system, and
$\mathbf{J}_{12}=\mathbf{J}_{1}+\mathbf{J}_{2}$ their sum. The
corresponding GK conductivities, $\kappa_{1}$, $\kappa_{2}$, and
$\kappa_{12}$ satisfy the relation:
$\left|\kappa_{12}-\kappa_{1}-\kappa_{2}\right|\le2\sqrt{\kappa_{1}\kappa_{2}}$.

\emph{Proof}---Let the \emph{energy displacement} associated with the
flux $\mathbf{J}_{i}$ be defined as:
$\mathbf{D}_{i}(t)=\frac{1}{\sqrt{6Vk_{B}T^{2}}}\int_{0}^{t}\mathbf{J}_{i}(t')dt'$.
The standard Einstein relation \cite{Helfand:1960} states that:
$\kappa_{i}=\lim_{t\to\infty}\left\langle
  |\mathbf{D}_{i}(t)|^{2}\right\rangle /t$;
it follows that:
$\kappa_{12}=\kappa_{1}+\kappa_{2}+\lim_{t\to\infty}2\left\langle
  \mathbf{D}_{1}(t)\cdot\mathbf{D}_{2}(t)\right\rangle /t$.
Canonical averages of products of phase-space functions can be seen as
scalar products: the lemma then follows from the Cauchy-Schwartz
inequality, as applied to the last relation. \hfill \qedsymbol

The application of the GK methodology to multi-component fluids
requires some generalizations because the presence of multiple atomic
species and the existence of additional hydrodynamical modes (one
conserved number per atomic species) do not permit to identify the
velocity field with the mass-current density, its integral with the
total momentum, and the heat flux with the energy flux. In molecular
fluids, however, this identification can still be done because the
integral of the velocity field, while not a constant, is a
non-diffusive flux, thus not contributing to the heat conductivity. In
order to demonstrate this statement, we first define the fluxes
$\mathbf{J}_{AB}=n_{B}\mbox{\textbf{V}}_{A}-n_{A}\mathbf{V}_{B}$,
where $A$ and $B$ indicate any two atomic species, $n_{A}/n_{B}$ their
stoichiometric ratio, and $\mathbf{V}_{S}=\sum_{i\in S}\mathbf{v}_{i}$
is the sum of the velocities of all the atoms of a same species
$S$. The integral $\int_0^t\mathbf{J}_{AB}(t')dt'$ is equal to the sum
of the variations of all the $AB$ relative positions in a same
molecule, which is bound by the sum of the variations of all the $AB$
distances. $\mathbf{J}_{AB}$ is therefore a
non-diffusive flux.  We have $N(N-1)/2$ such non-diffusive fluxes,
$N$ being the number of species, of which only $N-1$ are linearly
independent; furthermore the flux $J_{M}=\sum_{S}M_{S}\mathbf{V}_{S}$
($M_{S}$ is the mass of the $S$-th atomic species) is the total
momentum, and is thus non-diffusive. We have therefore $N$ independent
linear combinations of the $\mathbf{V}_{S}$ fluxes that are
non-diffusive.  We conclude that all of them, as well as their sum,
$\mathbf{V}(t) = \sum_{S} \mathbf{V}_{S}(t) = \frac{1}{V} \int
\mathbf{v}(\mathbf{r},t)d\mathbf{r}$, are also non-diffusive.

In order to derive an expression for the macroscopic energy flux
appearing in the GK formula, Eq. \eqref{eq:Green-Kubo}, we first
multiply the continuity equation, Eq. \eqref{eq:continuity}, by
$\mathbf{r}$ and integrate by parts, to obtain the first moment of the
time derivative of the energy density:
\begin{equation}
  \mathbf{J}_{e}(t) = \int\dot{e}(\mathbf{r},t)
  \mathbf{r}d\mathbf{r}. \label{eq:macroscopic-current}
\end{equation} In periodic boundary conditions (PBC)
Eq. \eqref{eq:macroscopic-current} is ill-defined for the very same
reason why macroscopic polarization in dielectrics is so 
\cite{Resta-Vanderbilt:2007}. In CMD the usual expression for the
energy flux in terms of atomic energies and forces
\cite{Allen-Tildesley} is recovered from
Eq. \eqref{eq:macroscopic-current} by the somewhat arbitrary
definition: 
$e(\mathbf{r},t) = \sum_{I}e_{I}(\mathbf{R},\mathbf{V})
\delta(\mathbf{r}-\mathbf{R}_{I})$,
where $e_{I}=\frac{1}{2}M_IV_I^2+\frac{1}{2}\sum_{J\ne I}v(|\mathbf{R}_J-\mathbf{R}_I|)$,
$\mathbf{R}=\{\mathbf{R}_{I}\}$, and $\mathbf{V}=\{\mathbf{V}_{I}\}$
are the atomic energies, positions, and velocities, and by reducing
the resulting expression to a boundary-insensitive form. In DFT an
energy density can be defined, which is however inherently
ill-determined because of the non-uniqueness of the
quantum-mechanical kinetic and classical electrostatic energy
densities \cite{Chetty:1992,Ramprasad:2002}. Our previous analysis
demonstrates that, in spite of previous worries to the contrary, the
transport coefficients derived from a DFT energy density through the
GK formula, Eq. \eqref{eq:Green-Kubo}, are well defined, provided a
macroscopic energy flux can be computed from
Eq. \eqref{eq:macroscopic-current} in PBC. To this end, among many
equivalent gauges, we choose to represent the DFT total energy as
the integral of the density:
\begin{widetext}
\begin{equation}
  e_{DFT}(\mathbf{r}) = \sum_{I}\delta(\mathbf{r}-\mathbf{R}_{I})
  e^0_I +\mathrm{Re}
  \sum_{v}\varphi_{v}^{*}(\mathbf{r}) \bigl(\hat{H}_{KS}
  \varphi_{v}(\mathbf{r})\bigr) -\frac{1}{2}
  \rho(\mathbf{r})v_{H}(\mathbf{r})
  +\left(\epsilon_{XC}(\mathbf{r})-v_{XC}(\mathbf{r})\right) 
  \rho(\mathbf{r}),\label{eq:epsilon_DFT}
\end{equation}
\end{widetext}
where $e^0_I=\frac{1}{2}M_{I}V_{I}^{2}+w_{I} $ are bare ionic
energies; $M_{I}$, $Z_{I}$, and
$w_{I}=\frac{1}{2}\sum_{J\ne
  I}\frac{Z_{I}Z_{J}}{|\mathbf{R}_{I}-\mathbf{R}_{J}|}$
being ionic masses, charges, and electrostatic energies, respectively;
the electron charge is assumed to be one; $\hat{H}_{KS}$ is the
instantaneous Kohn-Sham (KS) Hamiltonian, $\varphi_{v}$'s its occupied
eigenfunctions, and
$\rho(\mathbf{r}) = \sum_{v}| \varphi_{v}(\mathbf{r})|^{2}$ the
ground-state electron-density distribution; $v_{H}$ and $v_{XC}$ are
Hartree and exchange-correlation (XC) potentials, and $\epsilon_{XC}$
is a local XC energy per particle, defined by the relation:
$E_{XC}= \int \epsilon_{XC}
[\rho](\mathbf{r})\rho(\mathbf{r})d\mathbf{r}$.\footnote{$\epsilon_{XC}$
  is also to some extent ill-defined, in that any XC densities resulting
  in a same integral should be considered as equivalent.} The energy
density of Eq. \eqref{eq:epsilon_DFT} depends on time through
atomic positions and velocities and KS orbitals. Inserting its time
derivative into Eq. \eqref{eq:macroscopic-current} and using the
Born-Oppenheimer (BO) equations of motion for the nuclei
($M_{I}\mathbf{\dot{V}}_{I}=-\partial E_{DFT}/ \partial
\mathbf{R}_{I}$),
the resulting adiabatic energy flux can be expressed as:
\begin{equation}
  \mathbf{J}_{\epsilon}=\mathbf{J}_{KS} + \mathbf{J}_{H} +
  \mathbf{J}'_{0} + \mathbf{J}_{0}+\mathbf{J}_{XC}. \label{eq:5Js}
\end{equation}
The five fluxes in Eq. \eqref{eq:5Js} are defined as:
\begin{align}
  \mathbf{J}_{KS} 
  & =\sum_{v}
    \left (
    \langle\varphi_{v}| \mathbf{r}\hat{H}_{KS}|
    \dot{\varphi}_{v}\rangle + 
    \varepsilon_v \langle\dot{\varphi}_{v}|
                    \mathbf{r} |
    \varphi_{v}\rangle\right), \label{eq:J_KS}  \\
  \mathbf{J}_{H} & =\frac{1}{4\pi} \int\dot{v}_{H}(\mathbf{r})
                   \nabla v_{H}(\mathbf{r}) d\mathbf{r},\label{eq:J_H}\\
  \mathbf{J}'_{0} & =\sum_{v,I} \left\langle \varphi_{v}
                   \left|(\mathbf{r}-\mathbf{R}_{I}) \left(\mathbf{V}_{I}
                   \cdot
                   \nabla_I\hat v_0
                   \right)\right|\varphi_{v} \right\rangle ,\label{eq:J_0}\\
  \mathbf{J}_{0} & =\sum_{I}\Bigl [ \mathbf{V}_{I} e^0_I
                   + \sum_{L\ne I}
                   (\mathbf{R}_{I}- \mathbf{R}_{L})
                   \left(\mathbf{V}_{L} \cdot
                   \nabla_L w_I
                   \right)\Bigr ], \label{eq:J_n} \\
  \mathbf{J}_{XC} &=\begin{cases} 0 & \text{(LDA)} \\
      -\int\rho(\mathbf{r})\dot{\rho}(\mathbf{r}) \partial\epsilon_{GGA}
      (\mathbf{r})d\mathbf{r} & \text{(GGA)}, \label{eq:J_XC} \end{cases}
\end{align}
where $\varepsilon_v$ in Eq. \eqref{eq:J_KS} is the $v$-th eigenvalue
of the KS Hamiltonian; $\nabla=\frac{\partial}{\partial\mathbf{r}}$
and $\nabla_I=\frac{\partial}{\partial\mathbf{R}_I}$ in
Eqs. (\ref{eq:J_H}-\ref{eq:J_n}) indicate the gradients with respect
to the argument of the function and to the $I$-th atomic position,
respectively; the simbol $\hat v_0$ in Eq. \eqref{eq:J_0} indicates
the (possibly non-local) ionic (pseudo-) potential acting on the
electrons; finally, ``LDA'' and ``GGA'' in Eq. \eqref{eq:J_XC}
indicate the local-density \cite{KS:1965} and generalized-gradient
\cite{Perdew:1996} approximations to the XC energy functional,
respectively, and $\partial\epsilon_{GGA}$ the derivative of the GGA
XC local energy per particle with respect to density gradients.
Eq. \eqref{eq:5Js} can be derived from
Eqs. (\ref{eq:macroscopic-current}-\ref{eq:epsilon_DFT}) with some
tedious but straightforward algebra (see \emph{Methods}). The
last four terms on its right-hand side,
Eqs. (\ref{eq:J_H}--\ref{eq:J_XC}), are manifestly
boundary-insensitive, whilst the first, Eq. \eqref{eq:J_KS}, is not,
because the position operator appearing therein is ill-defined in
PBC. Within the adiabatic time evolution that is assumed in AIMD,
however, the time derivative of a KS orbital, as well as its product
with the KS Hamiltonian, are orthogonal to the orbital itself in the
``parallel transport'' gauge where KS orbitals are real
\cite{Thouless:1983,Resta:2000}:\footnote{The concept of \emph{gauge}
  for the quantum-mechanical representation of molecular orbitals
  should not be confused with that introduced in this paper for the
  energy density.}  $\langle\varphi_{v}|\dot{\varphi}_{v}\rangle=0$
and $\langle\varphi_{v}|H_{KS}|\dot{\varphi}_{v}\rangle=0.$ Therefore,
in order to evaluate Eq. \eqref{eq:J_KS}, one only needs the
projection of $\mathbf{r}|\varphi_{v}\rangle$ onto the manifold
orthogonal to \textbf{$\varphi_{v}$}, which is well defined in
PBC. Actually, by expanding $\dot{\varphi}_{v}$ in the basis of the
eigenstates of the instantaneous KS Hamiltonian \cite{Thouless:1983},
one sees that only the projection of $\mathbf{r}|\varphi_{v}\rangle$
onto the empty-state manifold,
$|\bar{\varphi}_{v}^{\alpha}\rangle=\hat{P}_{c}x^{\alpha}|\varphi_{v}\rangle,$
contributes to $\mathbf{J}_{KS}$, where
$\hat{P}_{c}=1-\sum_{v}|\varphi_{v}\rangle\langle\varphi_{v}|$ and
$x^{\alpha}$ is the $\alpha$-th Cartesian component of $\mathbf{r}$.
Using the standard prescription adopted in density-functional
perturbation theory (DFPT), such a projection can be computed by
solving the linear equation \cite{Baroni:2001}:
\begin{equation}
  (\hat{H}_{KS}-\varepsilon_v
  )|\bar{\varphi}_{v}^{\alpha} \rangle=
  \hat{P}_{c}[\hat{H}_{KS},x^{\alpha}]| \varphi_{v}\rangle,\label{eq:DFPT}
\end{equation}
where the ill-definedness of the solution, due to the singularity of
the left-hand side, is lifted by enforcing its orthogonality to the
occupied-state manifold.  In terms of the
$\bar{\varphi}_{v}^{\alpha}$'s Eq. \eqref{eq:J_KS} reads:
\begin{equation}
  J_{KS}^{\alpha} =\sum_{v}\left(\langle\bar{\varphi}_{v}^{\alpha}|
    H_{KS}|\dot{\varphi}_{v}\rangle +
    \varepsilon_v
    \langle\dot{\varphi}_{v}| \bar{\varphi}_{v}^{\alpha} \rangle\right).\label{eq:JKS}
\end{equation}

The flux in Eq. \eqref{eq:JKS} is not manifestly invariant with
respect to the arbitrary choice of the zero of the one-electron energy
levels. A shift of the energy zero by a quantity $\Delta\epsilon$
results in a shift of the Kohn-Sham energy flux:
$J_{KS}^{\alpha}\rightarrow J_{KS}^{\alpha} + \Delta\varepsilon
\sum_{v} \left(\langle\bar{\varphi}_{v}^{\alpha}|\dot{\varphi}_{v}
  \rangle+ \langle\dot{\varphi}_{v}|
  \bar{\varphi}_{v}^{\alpha}\rangle\right) =J_{KS}^{\alpha}+
\Delta\varepsilon J_{\rho}^{\alpha},$
where $\mathbf{J}_{\rho}$ is the adiabatic electronic macroscopic flux
introduced in Ref. \cite{Thouless:1983}. The electronic current is the
difference between the total charge current and its ionic component:
the first is by definition non-diffusive in insulators, while the
second is so in mono-atomic and molecular systems, as we have seen
when discussing the latter. We conclude that the electronic flux is
non-diffusive in insulators, thus not contributing to their heat
conductivity and lifting the apparent indeterminacy of
Eq. \eqref{eq:JKS}.

\section*{Numerical simulation}
The methodology presented above has been implemented in the
\textsc{Quantum ESPRESSO} suite of computer codes \cite{QE:2009}: a
Car-Parrinello (CP) \cite{Car:1985} AIMD trajectory is first generated
using the \texttt{cp.x} code; the energy flux is then evaluated along
this trajectory according to Eqs. (\ref{eq:5Js}-\ref{eq:J_n}) by an
add-on to the \texttt{pw.x} code implemented using several DFPT
routines borrowed from the \texttt{ph.x} code; the thermal
conductivity is finally computed from the GK relation,
Eq. \eqref{eq:Green-Kubo}, or the equivalent Einstein relation
\cite{Helfand:1960}.

In order to demonstrate this methodology, we compare its predictions
with those from CMD \cite{LAMMPS:1995} for a system whose DFT BO
energy surface can be accurately mimicked by pair potentials. Not
aiming at a realistic description of any specific system, but rather
at the ease and accuracy of the classical representation of the DFT BO
surface, we choose liquid Argon and use the LDA XC functional, in
spite of the well known inability of the latter to capture dispersion
forces. This reference system will be dubbed ``LDA-Ar''. KS
orbitals are treated within the plane-wave (PW) pseudo-potential (PP)
method \cite{PP}. Our model consists of 108 atoms in a periodically
repeated cubic supercell with an edge of 33 a.u., corresponding to a
density of 1.34 $\mathrm{g/cm^{3}}.$ AIMD trajectories were generated
via the Car-Parrinello dynamics \cite{Car:1985} for 100 picoseconds
(ps), using a time step of 0.242 femtoseconds (fs) and a fictitious
electronic mass of 1000 electronic masses, at two different
temperatures, $\mathrm{T}=250$ and $400\thinspace\mathrm{K}$. The
fictitious electronic temperature was monitored and checked not to be
subject to any significant drift. The BO energy surface was modeled
with a sum of classical pair potentials of the form
$V(r)=P_{2}(r)\mathrm{e}^{-\alpha r}$, where $P_{2}$ is a second-order
polynomial, whose parameters were determined independently for each
temperature by a least-square fit of the classical
vs. quantum-mechanical forces computed along the AIMD trajectory.
Self-diffusion coefficients of $(10.8\pm0.1)$, and
$(15.6\pm0.2)\times10^{-5}\mathrm{cm^{2}/s}$ were estimated along the
two AIMD trajectories, in close agreement with the CMD values
$(10.3\pm0.1)$, and $(15.8\pm0.2)\times10^{-5}\mathrm{cm^{2}/s}$, thus
confirming the quality of the classical model. Radial distribution
functions computed from AIMD and CMD trajectories were also found to
be very similar.

\begin{figure}[ht!]
\iftwocolumns\begin{centering}
\includegraphics[width=0.9\columnwidth]{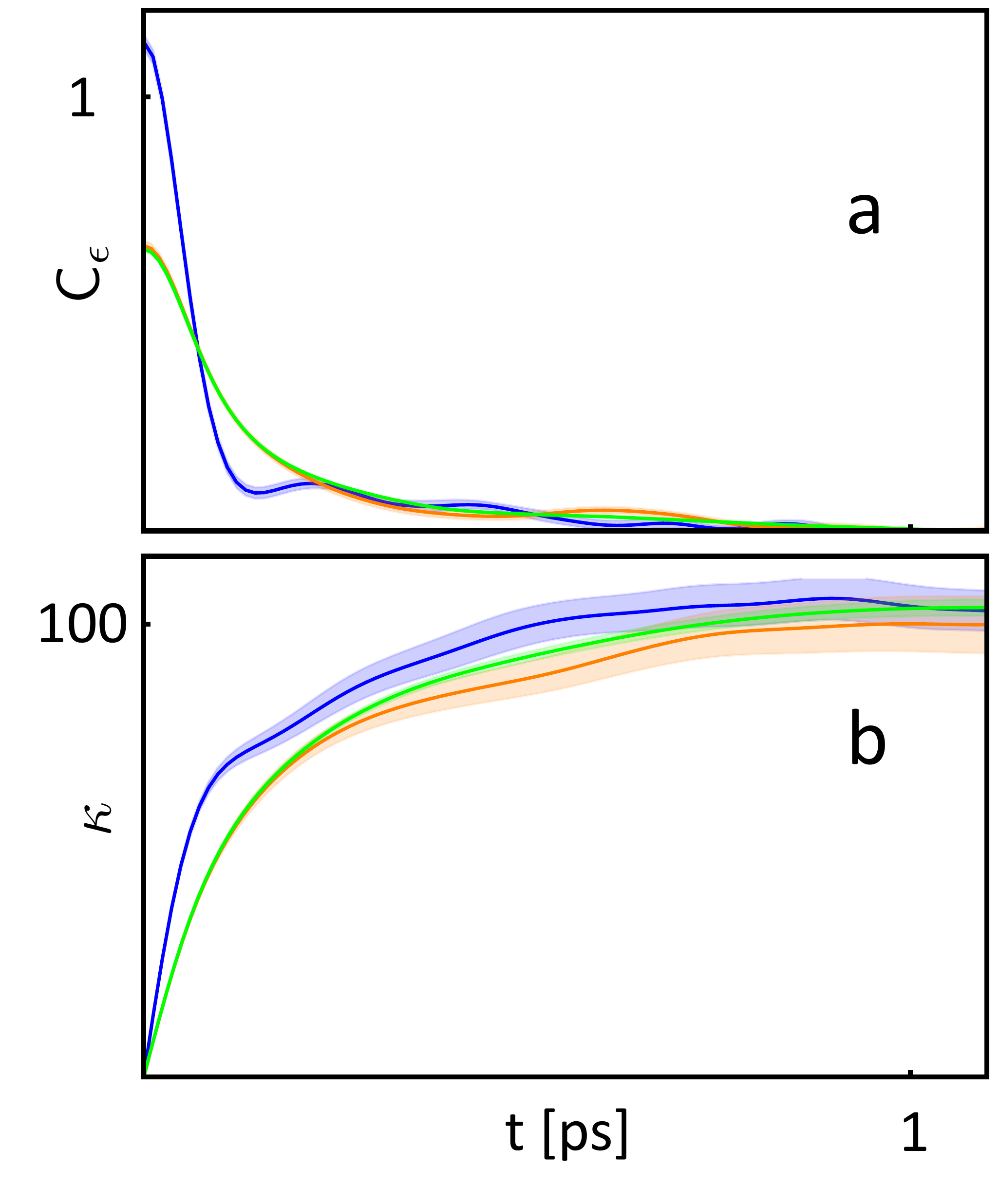}
\end{centering} \fi
\caption{Time correlations of the energy flux in LDA Ar. (a)
  $ C_{e}(t) = \frac{1}{3Vk_{B}T^{2}}
  \langle
  \mathbf{J}_{e}(t)\cdot\mathbf{J}_{e}(0)
  \rangle \thinspace
  \mathrm{[10^{15}mW K^{-1} s^{-1}]}$.
  (b) $\kappa(t)=\int_{0}^{t}C_{e}(t')dt'\thinspace
  \mathrm{[mW m^{-1}K^{-1}]}$
  (see text). Blue: AIMD (100 ps). Orange, CMD (100 ps). Green, CMD
  (1000 ps). The shaded areas depict statistical errors as
  estimated from a block analysis of our MD trajectories.
  \label{fig:Ar-correlation}
}
\end{figure}

In Fig. \ref{fig:Ar-correlation}a we compare the time correlation
functions of the energy flux in LDA Ar, as computed from AIMD and CMD
at $\mathrm{T=250\thinspace}\mathrm{K}$. The CMD and AIMD correlation
functions differ not quite because they correspond to different
systems---which are actually close enough as to have very similar
equilibrium and diffusion properties---as because the AIMD and CMD
fluxes derive from a different unpacking of the total energy into
local contributions.  In Fig. \ref{fig:Ar-correlation}b we display the
integrals
$\kappa(t) = \int_{0}^{t}\frac{1}{3Vk_{B}T^{2}}
\langle\mathbf{J}_{e}(t) \cdot\mathbf{J}_{e}(0)\rangle dt'$;
the AIMD and CMD heat conductivities,
$\kappa=\lim_{t\to\infty}\kappa(t)$, coincide within statistical
errors with each other and with the CMD value evaluated from a
1-ns-long simulation: ($103\pm5$, $100\pm6$, and $104\pm2$)
$[\mathrm{mW~K^{-1}m^{-1}}]$, respectively. A similar level of
agreement is obtained for the other temperature,
$\mathrm{T=400\thinspace}\mathrm{K}$ ($118\pm8$, $112\pm7$, and
$110\pm2$) $[\mathrm{mW~K^{-1}m^{-1}}]$.

\begin{figure}[ht!]
\iftwocolumns\begin{centering}
\includegraphics[width=0.9\columnwidth]{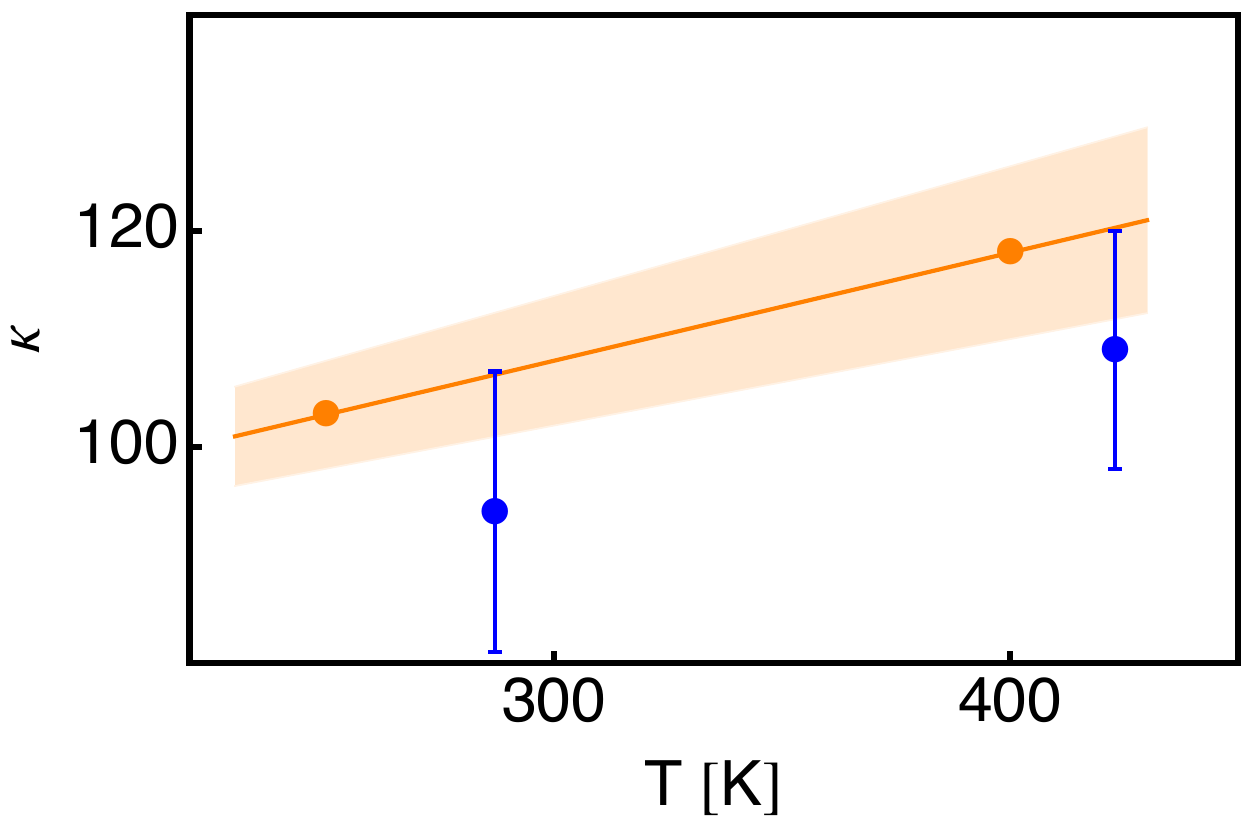}
\end{centering}\fi
\caption{ Comparison of the heat conductivities of our LDA-Ar model,
  as estimated from Green-Kubo and M\"uller-Plathe ab-initio molecular
  dynamics. Units are [$\mathrm{mW~K^{-1}m^{-1}}$]. Orange:
  equilibrium (GK) molecular dynamics; the two dots indicate the
  estimates from our simulations, the straight line their linear
  inter-/extrapolation. Statistical errors, as estimated by a block
  analysis of our MD trajectories, are indicated by error bars or by
  shaded areas, where relevant. \label{fig:MP-GK_comparison} }
\end{figure}

In order to further validate these results, we have recomputed the
thermal conductivities of our LDA-Argon model, using non-equilibrium
(MP) AIMD \cite{Mueller-Plathe:1997}. A detailed comparison of GK
\emph{vs.}  MP AIMD for heat-transport simulations is out of the scope
of the present paper, and we have limited ourselves to two MP
simulations, aimed at mimicking the physical conditions of the GK AIMD
simulations reported above, and performed using minimal simulation
settings: we used $( 2\times 2\times 5)$ supercells, where the
notation indicates multiples of a 4-atom cubic unit cell, thus
resulting in 80-atom tetragonal supercells whose size was chosen so as
to result in the same mass density of $1.34\thinspace \mathrm{g/cm^3}$
as used before. MP simulations were performed by subdividing the
supercell in eight equally spaced layers stacked along the c axis and
by swapping the velocities of the hottest atom in the cool region and
the coolest atom in the hot region every picosecond. Rather long
simulations ($\gtrsim 360~ \mathrm{ps}$) were necessary to achieve an
acceptable statistical accuracy, resulting in estimated thermal
conductivities of $94 \pm 13 $ and
$ 109 \pm 11 ~ [\mathrm{mW~K^{-1}m^{-1}}]$ at the temperatures of
$287$ and $423 ~ \mathrm{K}$, respectively. Our GK and MP AIMD results
are compared in Fig. \ref{fig:MP-GK_comparison}, witnessing to a
convincing validation of our approach based on the Green-Kubo
formalism.

\begin{figure}
\iftwocolumns\begin{centering}
\includegraphics[width=0.9\columnwidth]{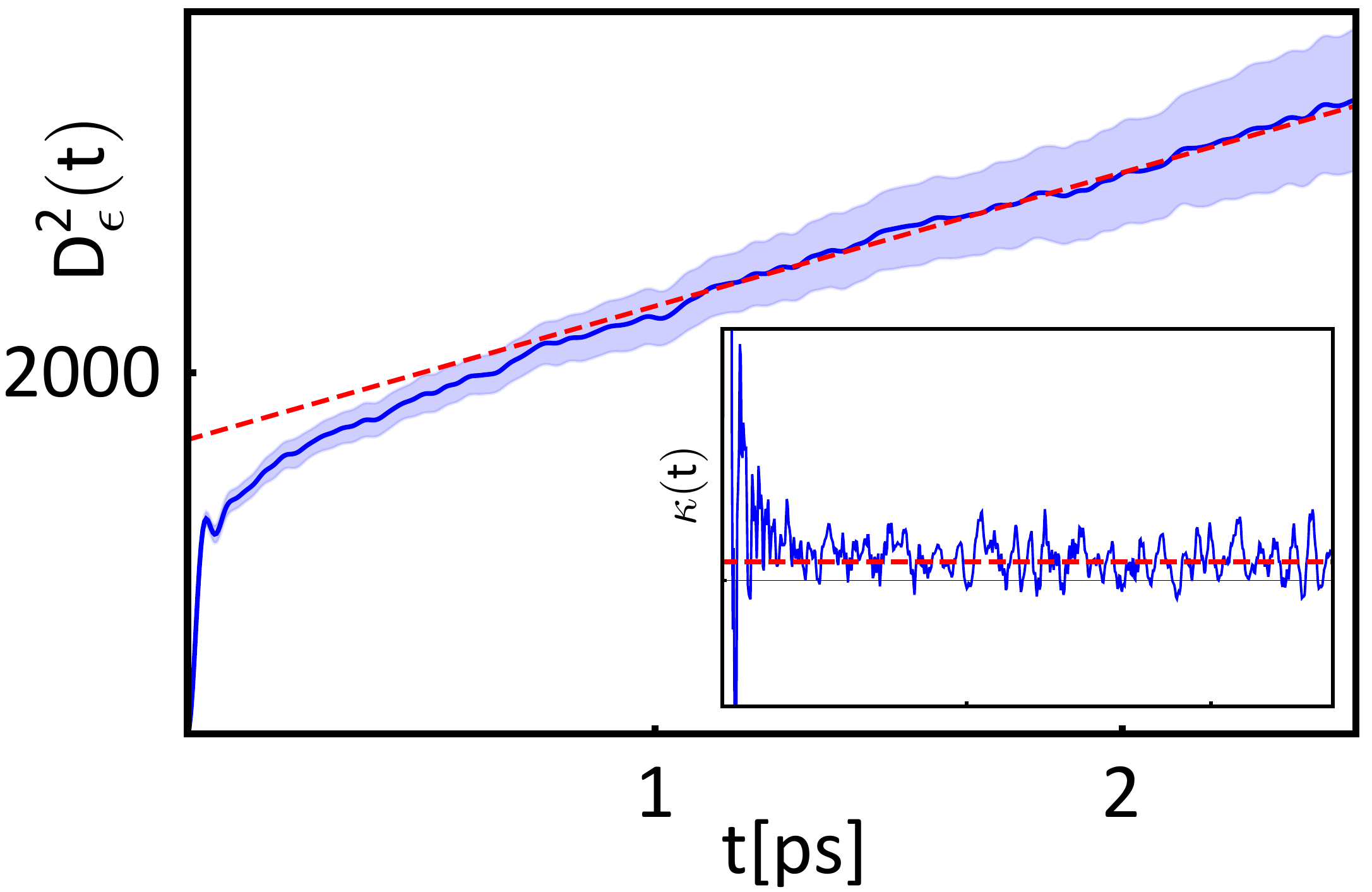}
\end{centering}\fi
\caption{Energy displacement in liquid heavy water at ambient
  conditions.
  $D_{e}^{2}(t) \thinspace \mathrm{[10^{-12}mJ m^{-1}K^{-1}]}$ was
  evaluated from the modified flux
  ${\mathbf{J}^*_{e}} =\mathbf{J}_{e}+\lambda^{*}\mathbf{V}$ at
  $T=\mathrm{385~K}$ with a GGA XC functional (see text); the dashed
  line indicates a linear fit to the large-time behavior of the
  curve. Inset: integral in the GK relation, $\kappa(t)$
  Eq. \eqref{eq:Green-Kubo}, as a function of the upper limit of
  integration (see caption to Fig. 1); the dashed line indicates the
  value of the thermal conductivity obtained from the Einstein
  relation. \emph{i.e. }from the slope of above linear fit. The shaded
  areas depict statistical errors, as estimated from a block analysis
  of our MD trajectories.\label{fig:H2O-Einstein}}
\end{figure}

We have applied our newly developed method to compute the heat
conductiviy of heavy water at ambient conditions. We have generated a
90-ps long AIMD trajectory for a system of 64 heavy-water molecules in
a cubic supercell with an edge of 23.46 a.u., corresponding to the
experimental density of 1.11 $\mathrm{g/cm^{3}}$, and at an estimated
temperature $\mathrm{T=385\thinspace K}$, using the PBE XC energy
functional \cite{Perdew:1996} and the PW-PP method as above
\cite{PP}. A time step of 0.0726 fs and a fictitious electronic mass
of 340 electron masses were used in this case. The resulting
self-diffusion coefficient was estimated to
$(2.6\pm0.2)\times10^{-5}\thinspace\mathrm{cm^{2}/s}$, to be compared
with an experimental value of
$2.0\times10^{-5}\thinspace\mathrm{cm^{2}/s}$ at
$T=298\mathrm{\thinspace K}$, following the common practice of
comparing experimental data for water at ambient conditions with
AIMD-PBE simulations performed at $ \sim400\:\mathrm{K}$
\cite{Grossman2004}. The power spectrum of the computed energy flux is
characterized by three relatively narrow peaks in correspondence to
the intramolecular vibrational modes \cite{Silvestrelli:1997},
resulting in long-lived high-frequency oscillations in the integrand
of Eq. \eqref{eq:Green-Kubo}, that plague the evaluation of the
integral as a function of the upper limit of integration well beyond
the time where the noise of the integrand becomes larger than the
amplitude of its oscillations. As the computation of transport
coefficients from the Einstein relation \cite{Helfand:1960} is less
affected by the high-frequency components of the power spectrum
\cite{Marcolongo:2014}, this ailment is alleviated by evaluating the
heat conductivity as the slope of the \emph{energy squared
  displacement},
$D_{e}^{2}(t)=\frac{1}{6Vk_{B}T^{2}}\left\langle
  \left|\int_{0}^{t}\mathbf{J}_{e}(t')dt'\right|^{2}\right\rangle $,
as a function of $t$ in the large-time limit. A direct application of
this technique is however not possible as the long-time behavior of
the energy squared displacement does not allow us to extrapolate a
straight line before it becomes too noisy to analyze. This state of
affairs indicates the existence of a slowly decaying mode in the
energy-flux correlation function, possibly correlated with a
non-diffusive flux. As we have seen, the total velocity $\mathbf{V}$
is such a non-diffusive flux. The value of the corresponding GK
conductivity, Eq. \eqref{eq:Green-Kubo}, however, goes to zero very
slowly as a function of the upper limit of integration. This suggests
that the slow convergence of the heat conductivity of water as
estimated from the slope of the energy squared displacament as a
function of time, is possibly due to large correlations existing
between the energy flux and the total velocity. We have therefore
decided to analyze, instead of $\mathbf{J}_{e}$, the modified flux
$\mathbf{J}^*_{e}= \mathbf{J}_{e} + \lambda^* \mathbf{V}$, where
$\lambda^{*}$ has been fixed in such a way as to minimize the
correlations between $\mathbf{J}{}_{e}^{*}$ and $\mathbf{V}$.  Fig. 2
displays the squared energy displacement computed from
$\mathbf{J}_{e}^{*}$ as a function of time and demonstrates that a
constant slope can indeed be identified in the long-time limit, giving
a value for the heat conductivity of heavy water of
$740\pm 120\thinspace \mathrm{mW m^{-1} K^{-1}}$, to be compared
with an experimental value of
$606 \thinspace \mathrm{mW m^{-1} K^{-1}}$ and
$595 \thinspace \mathrm{mW m^{-1} K^{-1}}$ for light and heavy
water respectively at ambient conditions \cite{Expt:H2O}.  The inset
displays the behavior of $\kappa(t)$ (see caption to Fig.  1) as a
function of the upper limit of integration in the GK formula,
indicating that a direct use of Eq. \eqref{eq:Green-Kubo} would be
extremely difficult in this case. A more detailed error analysis and a
systematic extension of this study to different isotopic compositions
and other conditions of temperature and pressure is currently in the
works.

\section*{Conclusions}
We believe that the discussion presented in this work will elucidate
the scope of a number of assumptions that, although routinely made in
the classical simulation of heat transport, have never been fully
clarified, thus hampering their generalization to quantum simulations.
We are confident that the resulting new methodology will have an
impact on important problems where other methods may fail, such as
\emph{e.g.} liquids and glasses, particularly at extreme conditions of
temperature and pressure.

\section*{Methods}

In order to derive Eqs. (\ref{eq:5Js}--\ref{eq:J_XC}), we start from
Eq. \eqref{eq:epsilon_DFT}, which we rewrite as:
\begin{equation}
  e_{DFT}(\mathbf{r}) =
  e_{KS}(\mathbf{r}) +
  e_{0}(\mathbf{r}) +  
  e_{H}(\mathbf{r}) +  
  e_{XC}(\mathbf{r}), \label{eq:esplit}
\end{equation}
where
\begin{align}
  e_{KS}(\mathbf{r}) 
  &= \mathrm{Re} \sum_{v}\varphi_{v}^{*}(\mathbf{r}) \bigl(\hat{H}_{KS}
    \varphi_{v}(\mathbf{r})\bigr), \label{eq:epsilon_KS}\\
  e_{0}(\mathbf{r}) 
  &= \sum_{I}\delta(\mathbf{r}-\mathbf{R}_{I})
    \left(\frac{1}{2}M_{I}V_{I}^{2}+w_{I}\right), \label{eq:epsilon_n}\\ 
  e_{H}(\mathbf{r}) 
  &= -\frac{1}{2} \rho(\mathbf{r})v_{H}(\mathbf{r}), \qquad\qquad\qquad
    \text{and} \label{eq:epsilon_H}  \\
  e_{XC}(\mathbf{r}) 
  &= \left(\epsilon_{XC}(\mathbf{r})-v_{XC}(\mathbf{r})\right)
    \rho(\mathbf{r}), \label{eq:epsilon_XC}
\end{align}
$\epsilon_{XC}$ is a local XC energy per particle, defined by
the relation
\begin{equation} E_{XC}= \int \epsilon_{XC}
[\rho](\mathbf{r})\rho(\mathbf{r})d\mathbf{r}, \label{eq:EXC}
\end{equation}
and the XC potential $v_{XC}$ is
\begin{align}
  v_{XC}(\mathbf{r}) 
  &= \frac{\delta E_{XC}}{\delta \rho(\mathbf{r})} \nonumber \\ 
  &= \epsilon_{XC}(\mathbf{r}) + \int \frac{\delta \epsilon_{XC}
    (\mathbf{r}')} {\delta\rho(\mathbf{r})} \rho(\mathbf{r}')
    d\mathbf{r}'. \label{eq:vXC}
\end{align}
In the LDA, $\epsilon_{XC}$ is a function of the local density,
whereas in the GGA it is a function of the local density and density
gradients:
\begin{align}
  \epsilon_{XC}^{LDA}[\rho](\mathbf{r})
  & =\epsilon_{LDA}\bigl (\rho(\mathbf{r})\bigr ), \\
  \epsilon^{GGA}_{XC}[\rho](\mathbf{r})
  & =\epsilon_{GGA}\bigl (\rho(\mathbf{r}),\nabla\rho(\mathbf{r})
    \bigr ). \label{eq:GGA}
\end{align}

We now proceed to computing the first moments of the
time derivatives of the above four densities, according to
Eq. \eqref{eq:macroscopic-current}. In order to simplify the notation,
the time dependence of the various quantities
will be omitted.
Let's start with the Kohn-Sham
energy density, Eq. \eqref{eq:epsilon_KS}.
\begin{widetext}
  \begin{align}
    \dot{e}_{KS}(\mathbf{r}) 
      & =\sum_{v} \left( \dot{\varphi}_{v}^{*}(\mathbf{r}) \hat{H}_{KS}
        \varphi_{v}(\mathbf{r})+ \varphi_{v}^{*}(\mathbf{r}) \hat{H}_{KS}
        \dot{\varphi}_{v}(\mathbf{r}) + \varphi_{v}^{*}(\mathbf{r})
        \dot{\hat{H}}_{KS}  \varphi_{v}(\mathbf{r})
        \right) \label{eq:epsilon_dot_KS_1} \\
      & =\dot{\bar{e}}_{KS}(\mathbf{r}) + \dot{e}'_{0}
        (\mathbf{r})+ \dot{e}_{H}'(\mathbf{r}) +
        \dot{e}'_{XC}(\mathbf{r}), \label{eq:epsilon_dot_KS_2} 
  \end{align}
\end{widetext}
where
\begin{align}
  \dot{\bar{e}}_{KS}(\mathbf{r})
  & =\sum_{v} \left[ \varepsilon_v\dot{\varphi}_{v}^{*}(\mathbf{r})
    \varphi_{v}(\mathbf{r})
    +\varphi_{v}^{*}(\mathbf{r})\hat{H}_{KS}
    \dot{\varphi}_{v}(\mathbf{r}) \right], \label{eq:dot_bar_epsilon_KS}\\
  \dot{e}'_{0}(\mathbf{r})
  & =\sum_{v} \varphi_{v}^{*}(\mathbf{r})
    \dot{\hat{v}}_{0}
    \varphi_{n}(\mathbf{r}), \label{eq:dot_epsilon_0}  \\
  \dot{e}_{H}'(\mathbf{r})
  &=\dot{v}_{H}(\mathbf{r})\rho(\mathbf{r}),\qquad
    \text{and} \label{eq:dot_epsilon_prime_H}\\ 
  \dot{e}'_{XC}(\mathbf{r})
  & = \dot{v}_{XC}(\mathbf{r})
    \rho(\mathbf{r}). \label{eq:dot_epsilon_prime_XC}  
\end{align}
The macrosopic flux deriving from $\dot{\bar{e}}_{KS}$,
Eq. \eqref{eq:dot_bar_epsilon_KS}, is the ``Kohn-Sham'' flux of
Eq. \eqref{eq:J_KS}:
\begin{equation}
  \int\mathbf{r}
   \dot{\bar{e}}_{KS}(\mathbf{r})  d\mathbf{r} =
  \mathbf{J}_{KS}.
\end{equation}
The other three terms,
Eqs. (\ref{eq:dot_epsilon_0}--\ref{eq:dot_epsilon_prime_XC}) result
from the external-, Hartree-, and XC-potential contributions to the
time derivative of the KS Hamiltonian (third term in
Eq. \ref{eq:epsilon_dot_KS_1}).  The corresponding fluxes combine with
the fluxes originating from the energy densities of
Eqs. (\ref{eq:epsilon_n}--\ref{eq:epsilon_XC}), as explained below.

The first moment of the ``ionic potential'' energy-density derivative,
Eq. \eqref{eq:dot_epsilon_0}, reads:
\begin{widetext}
\begin{align}
\int\mathbf{r}\dot{e}'_{0}(\mathbf{r})d\mathbf{r}
  & =\sum_{v}\langle\varphi_{v}| \mathbf{r} \dot{\hat{v}}_{0}|
    \varphi_{v}\rangle \nonumber \\
  & =\sum_{v,I} \left\langle \varphi_{v} \left| \mathbf{r}
    \left( \mathbf{V}_{I} \cdot \nabla_I \hat{v}_{0} \right) \right|
    \varphi_{v} \right\rangle \nonumber \\
  & = \sum_{v,I} \Bigl [ \left\langle \varphi_{n} \left| (\mathbf{r} - \mathbf{R}_I)
    \left( \mathbf{V}_{I} \cdot \nabla_I \hat{v}_{0} \right) \right|
    \varphi_{n} \right\rangle
+ \mathbf{R}_I \langle\varphi_{v}| ( \mathbf{V}_I
    \cdot \nabla_I{\hat{v}}_{0} ) |\varphi_{v}\rangle \Bigr ]
    \nonumber \\ 
  & =\mathbf{J}'_{0} - \sum_{I} \mathbf{R}_{I} \left( \mathbf{V}_{I}
    \cdot \mathbf{F}_{I}^{el} \right), \label{eq:J_0_prime}
 \end{align}
\end{widetext}
where $\mathbf{J}'_0$ is the flux of Eq. \eqref{eq:J_0}, and
$\mathbf{F}^{el}_I$ is the electronic (Hellmann-Feynman) contribution
to the force acting on the $I$-th atom. The corresponding (second)
term in the energy flux of Eq. \eqref{eq:J_0_prime}
is ill-defined in PBC but,
as we will see shortly, it cancels with a similar term coming from
the first moment of the ``ionic'' energy density, Eq. \eqref{eq:epsilon_n}.

The time derivative of the ``ionic'' energy density,
Eq. \eqref{eq:epsilon_n}, reads:
\begin{widetext}
\begin{equation}
  \dot e_{0}(\mathbf{r}) =
  \sum_{I} 
  \bigg [ 
  e^0_I \, \mathbf{V}_I\cdot \nabla_I\delta(\mathbf{r}
  -\mathbf{R}_{I}) +\delta(\mathbf{r} -\mathbf{R}_{I})
  \bigg(
  M_I \mathbf{V}_{I} \cdot\dot{\mathbf{V}}_{I} +
  \sum_{J\ne I} \mathbf{V}_{J} \cdot
  \nabla_J w_I
  \bigg)
  \bigg].
\end{equation}
\end{widetext}
We now use Newton's equations of motion
($M_{I} \dot{\mathbf{V}}_{I} = \mathbf{F}_I$, where $\mathbf{F}_I$ is
the force acting on the $I$-th atom), and split $ \mathbf{F}_I$ into
an electronic (Hellmann-Feynman) contribution, plus a sum of pair-wise
electrostatic terms,
$\mathbf{F}_I = \mathbf{F}^{el}_I-\sum_{J\ne I} \nabla_I w_J $, to
obtain:
\begin{widetext}
\begin{align}
  \int\mathbf{r}\dot{e}_{0}(\mathbf{r})d\mathbf{r}
  &=  \sum_I 
    \Bigl [
    e^0_I\mathbf{V}_I
    +
    \mathbf{R}_I 
    \Bigl (
    \mathbf{F}_I\cdot\mathbf{V}_I
    + \sum_{J\ne I} 
    \mathbf{V}_J\cdot
    \nabla_J w_I
    \Bigr ) 
    \Bigr ] . \nonumber \\
  & =\sum_I 
    \Bigl [
    e^0_I\mathbf{V}_I
    + \mathbf{R}_I 
    \left (
    \mathbf{F}^{el}_I\cdot\mathbf{V}_I 
    \right ) + \mathbf{R}_I \sum_{J\ne I} 
    \Bigl (
      \mathbf{V}_J\cdot 
      \nabla_J w_I
      - \mathbf{V}_I\cdot 
      \nabla_I w_J
    \Bigr ) 
  \Bigr ]  \nonumber \\
  & =\sum_I 
    \Bigl [
    e^0_I\mathbf{V}_I
    + \mathbf{R}_I 
    \left (
    \mathbf{F}^{el}_I\cdot\mathbf{V}_I 
    \right ) +\sum_{J\ne I} (\mathbf{R}_I-\mathbf{R}_J)(\mathbf{V}_J\cdot
    \nabla_J w_I) \bigr ] \nonumber \\
  &= \mathbf{J}_0+\sum_I \mathbf{R}_I 
    \left (
    \mathbf{F}^{el}_I\cdot\mathbf{V}_I 
    \right ), \label{eq:J_0_k}
\end{align}
\end{widetext}
where $\mathbf{J}_0$ is the energy flux of Eq. \eqref{eq:J_n} and the
third step follows from the second by interchanging the dummy indeces
of one of the two sums over $I$ and $J$. As anticipated before, the
second term on the right-hand side of Eq. \eqref{eq:J_0_k}, which is
ill-defined in PBC, cancels a similar term in
Eq. \eqref{eq:J_0_prime}, leaving all the surviving terms well
defined. We summarize Eqs.  \eqref{eq:J_0_prime} and \eqref{eq:J_0_k}
as:
\begin{equation}
  \int\mathbf{r}
  \bigl [ \dot{e}_{0}(\mathbf{r}) +
  \dot{e}'_{0}(\mathbf{r}) \bigr ]
  d\mathbf{r} =
  \mathbf{J}_0+\mathbf{J}'_0,
\end{equation}
where  $\mathbf{J}'_0$ and $\mathbf{J}_0$ are the energy fluxes of
Eqs. \eqref{eq:J_0} and \eqref{eq:J_n}, respectively.

We then combine the time derivative of the ``Hartree'' energy density,
Eq. \eqref{eq:epsilon_H}, with the ``Hartree-potential''
energy-density derivative, Eq. \eqref{eq:dot_epsilon_prime_H}:
\begin{align}
  \dot{\bar{e}}_H(\mathbf{r}) 
  &= \dot e_H(\mathbf{r})  + \dot e'_H(\mathbf{r}) \noindent \\
  & = \frac{1}{2} \bigl ( \dot v_H(\mathbf{r}) \rho(\mathbf{r}) -
    \dot\rho(\mathbf{r}) v_H(\mathbf{r})  \bigr ) \nonumber \\
  &=  \frac{1}{8\pi} \bigl ( v_H(\mathbf{r}) \Delta\dot
    v_H(\mathbf{r}) - \dot v_H(\mathbf{r}) \Delta v_H(\mathbf{r})
    \bigr ) \nonumber \\
  &=  \frac{1}{8\pi} \nabla \cdot \bigl ( v_H(\mathbf{r}) \nabla\dot
    v_H(\mathbf{r}) - \dot v_H(\mathbf{r}) \nabla v_H(\mathbf{r})
    \bigr ) \label{eq:dot_epsilon_comb}
\end{align}
Multiplying Eq. \eqref{eq:dot_epsilon_comb} by $\mathbf{r}$ and
integrating by parts, one obtains:
\begin{align}
  \mathbf{J}_H
  &= \int\mathbf{r} \dot{\bar{e}}_H(\mathbf{r}) d\mathbf{r} \nonumber \\
  &= \frac{1}{4\pi} \int \dot{v}_H(\mathbf{r}) \nabla {v}_H(\mathbf{r}) d\mathbf{r},
\end{align}
which is Eq. \eqref{eq:J_H}.

\begin{figure}[h]
\iftwocolumns\begin{centering}
\includegraphics[width=0.9\columnwidth]{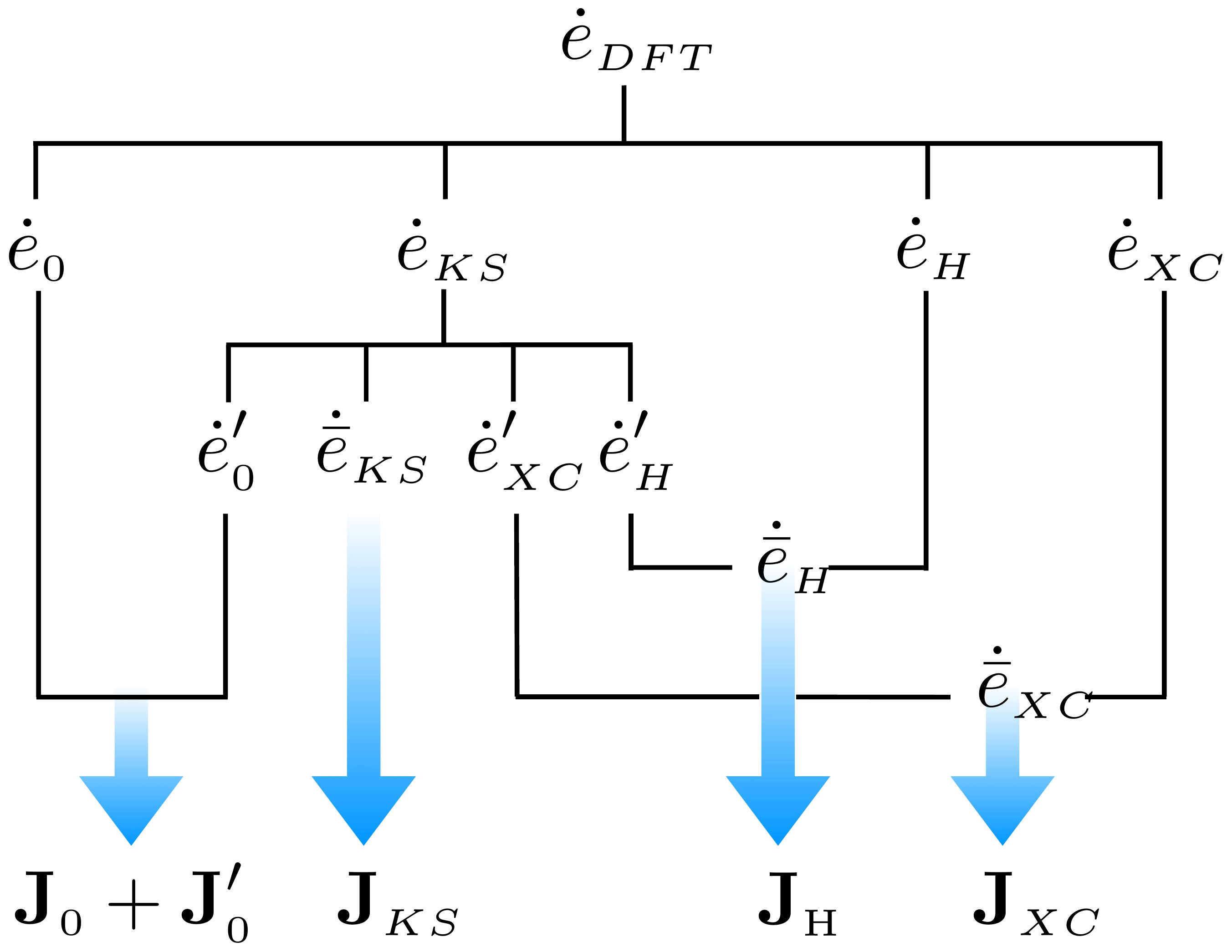}
\end{centering}\fi
\caption{
Conceptual flow of the derivation of the various components of the
macroscopic energy flux,
Eqs. (\ref{eq:J_KS}-\ref{eq:J_XC}), from the definition of a
microscopic energy density, Eqs. (\ref{eq:esplit}-\ref{eq:epsilon_XC})
  \label{fig:flow}
}
\end{figure}

We finally address the first moments of the time derivative of the ``XC'' energy density,
Eq. \eqref{eq:epsilon_XC}, and of the ``XC-potential'' energy-density derivative,
Eq. \eqref{eq:dot_epsilon_prime_XC}. We define:
\begin{widetext}
\begin{align}
\dot{\bar{e}}_{XC}(\mathbf{r})
  &=  \dot{e}_{XC}(\mathbf{r}) + \dot{e}'_{XC}(\mathbf{r}) \nonumber \\
  &= \bigl ( \epsilon_{XC}(\mathbf{r}) - v_{XC}(\mathbf{r}) \bigr )
    \dot\rho(\mathbf{r}) + \dot\epsilon_{XC}(\mathbf{r})
    \rho(\mathbf{r}) \nonumber \\
  &= \rho(\mathbf{r}) \int
    \frac{\delta\epsilon_{XC}(\mathbf{r})}{\delta\rho(\mathbf{r}')}
    \dot\rho(\mathbf{r}')d\mathbf{r}'
- \dot \rho(\mathbf{r}) \int
    \frac{\delta\epsilon_{XC}(\mathbf{r}')}{\delta\rho(\mathbf{r})}
    \rho(\mathbf{r}')d\mathbf{r}', \label{eq:barexc}
\end{align}
\end{widetext}
which derives from the definition of the XC potential, Eq. \eqref{eq:vXC},
and from the chain rule as applied to the time derivative of
$\epsilon_{XC}$: 
\begin{equation}
  \dot\epsilon_{XC}(\mathbf{r}) = \int
  \frac{\delta\epsilon_{XC}(\mathbf{r})}{\delta\rho(\mathbf{r}')
  }\dot\rho(\mathbf{r}')d\mathbf{r}'.
\end{equation}
The first moment of Eq. \eqref{eq:barexc} reads:
\begin{align}
  \mathbf{J}_{XC} &= \int\mathbf{r} \dot{\bar{e}}_{XC}(\mathbf{r})
                    d\mathbf{r} \nonumber \\ 
&= \int (\mathbf{r} - \mathbf{r}')  \rho(\mathbf{r})
  \dot\rho(\mathbf{r}') \frac{\delta\epsilon_{XC}(\mathbf{r})}{\delta
  \rho(\mathbf{r}')} d\mathbf{r} d\mathbf{r}'. \label{eq:JXCs}
\end{align}
In the LDA, because of the local dependence of $\epsilon_{XC}$ on the
electron density, the functional derivative in Eq. \eqref{eq:JXCs} is
proportional to $\delta(\mathbf{r}-\mathbf{r}')$, thus making the
integral vanish. In the GGA Eq. \eqref{eq:GGA} gives:
\iftwocolumns
\begin{multline}
  \frac{\delta\epsilon_{XC}^{GGA}(\mathbf{r})}{\delta \rho(\mathbf{r}')}
  = \epsilon'_{GGA}(\mathbf{r}) \delta(\mathbf{r}-\mathbf{r}') \\ +
  \sum_{\alpha}\partial_\alpha\epsilon_{GGA}(\mathbf{r})
  \nabla_{\alpha} \delta(\mathbf{r}-\mathbf{r}'), \label{eq:deltaGGA}
\end{multline}
\else
\begin{equation}
  \frac{\delta\epsilon_{XC}^{GGA}(\mathbf{r})}{\delta \rho(\mathbf{r}')}
  = \epsilon'_{GGA}(\mathbf{r}) \delta(\mathbf{r}-\mathbf{r}') \\ +
  \sum_{\alpha}\partial_\alpha\epsilon_{GGA}(\mathbf{r})
  \nabla_{\alpha} \delta(\mathbf{r}-\mathbf{r}'), \label{eq:deltaGGA}
\end{equation}
\fi
where
$\epsilon'_{GGA}(\mathbf{r})\doteq \left
  . \frac{\partial\epsilon_{GGA}(\rho,\nabla\rho)}{\partial\rho}
\right |_{\rho=\rho(\mathbf{r})}$,
and
$\partial_\alpha\epsilon_{GGA}(\mathbf{r})\doteq \left
  . \frac{\partial\epsilon_{GGA}(\rho,\nabla\rho)}{\partial\nabla_\alpha\rho}
\right |_{\rho=\rho(\mathbf{r})}$.
The first term on the right-hand side of Eq. \eqref{eq:deltaGGA} does
not contribute to the XC energy flux as in the LDA. By inserting the
second term into Eq. \eqref{eq:JXCs}, one finally arrives at the
expression for the XC energy flux of Eq. \eqref{eq:J_XC}, thus
completing the derivation of
Eqs. (\ref{eq:J_KS}-\ref{eq:J_XC}). This rather unwieldy, but all
in all straightfoward, derivation is visually summarized in
Fig. \ref{fig:flow}.

\iftwocolumns\else
\section*{Corresponding author}
All correspondence should be addressed to SB, \texttt{baroni@sissa.it}
\fi

\section*{Acknowledgements}
SB gratefully acknowledges useful discussions with Tao Sun and Dario
Alf\`e in the early phases of this work and, more recently, with
Roberto Car and Raffaele Resta. All the authors gratefully thank
Luciano Colombo, Claudio Melis, Simon R. Philpot, and Aleksandr Chernatynskiy
for commucating to them some of their unpublished material.
 
\section*{Author contributions}
All authors contributed to all aspects of this work.
\iftwocolumns\else\newpage\fi


\begin{thebibliography}{10}

\bibitem{Klemens:1959} P.\,G. Klemens, \emph{Thermal conductivity and
    lattice vibrational modes}, Solid State Phys. \textbf{7},
  1\textendash 98 (1958).

\bibitem{Green:1954} M.\,S. Green, \emph{Markoff random processes and the statistical mechanics of time-dependent phenomena, II.
Irreversible processes in fluids}, J. Chem. Phys. \textbf{22,}
  398\textendash 413 (1954).

\bibitem{Kubo} R. Kubo, \emph{Statistical-mechanical theory of
    irreversible processes. I.  General Theory and Simple Applications
    to Magnetic and Conduction Problems},
  J. Phys. Soc. Jpn. \textbf{12}, 570\textendash586 (1957).

\bibitem{Evans-Morriss} D.J. Evans and G. Morriss, \emph{Statistical
    mechanics of nonequilibrium liquids} 2nd ed. (Cambridge University
  Press, Cambridge UK, 2008).

\bibitem{Allen-Tildesley} M.\,P. Allen and D.\,J. Tildesley,
  \emph{Computer simulation of liquids} (Clarendon Press, Oxford,
  1987).

\bibitem{Mueller-Plathe:1997} F. M\" uller-Plather, \emph{A simple
    nonequilibrium molecular dynamics method for calculating the
    thermal conductivity}, J. Chem. Phys. \textbf{106},
  6082\textendash6085 (1997).

\bibitem{Car:1985} R. Car and M. Parrinello, \emph{Unified approach
    for molecular dynamics and density functional theory},
  Phys. Rev. Lett. \textbf{55}, 2471\textendash2474 (1985).

\bibitem{Marx-Hutter} D. Marx and J. Hutter, \emph{Ab initio molecular
    dynamics} (Cambridge University Press, Cambridge UK, 2012).

\bibitem{Stackhouse:2010} S. Stackhouse, L. Stixrude, and B.\,B.
  Karki, \emph{Thermal conductivitity of periclase (MgO) from first
    principles}, Phys. Rev. Lett. \textbf{104}, 208501 (2010).

\bibitem{HK:1964} P. Hohenberg and W. Kohn, \emph{Inhomogeneous
    electron gas}, Phys. Rev. \textbf{136}, B864\textendash B871
  (1964).

\bibitem{KS:1965} W. Kohn and L. Sham, \emph{Self-consistent equations
    including exchange and correlation effects},
  Phys. Rev. \textbf{140}, A1133\textendash A1138 (1965).

\bibitem{Kadanoff-Martin} L.\,P. Kadanoff and P.\,C. Martin,
  \emph{Hydrodynamic equations and correlation functions},
  Ann.Phys. \textbf{24}, 419\textendash 469 (1963).

\bibitem{Forster} D. Forster, \emph{Hydrodynamic fluctuations, broken
    symmetry, and correlation functions} (Benjamin, Reading, 1975).

\bibitem{Helfand:1960} E. Helfand, \emph{Transport coefficients from
    dissipation in a canonical ensemble}, Phys. Rev. \textbf{119},
  1\textendash 9 (1960).

\bibitem{Resta-Vanderbilt:2007} R. Resta and D. Vanderbilt,
  \emph{Theory of polarization: a modern approach},
  Top. Appl. Phys. \textbf{105}, 31\textendash 68 (2007).

\bibitem{Chetty:1992} N. Chetty and R.\,M. Martin,
  \emph{First-principles energy density and its applications to
    selected polar surfaces}, Phys. Rev. B \textbf{45},
  6074\textendash 6088 (1992).

\bibitem{Ramprasad:2002} R. Ramprasad, \emph{First-principles energy
    and stress fields in defected materials}, J. Phys.:
  Condens. Matter, \textbf{14}, 5497\textendash 5516 (2002).

\bibitem{Perdew:1996} J.\,P. Perdew, K. Burke, and M. Ernzerhof,
  \emph{Generalized gradient approximation made simple},
  Phys. Rev. Lett. \textbf{77}, 3865\textendash 3868 (1996).

\bibitem{Thouless:1983} D.J. Thouless, \emph{Quantization of particle
    transport}, Phys. Rev. B \textbf{27}, 6083\textendash 6087 (1983).

\bibitem{Resta:2000} R. Resta, \emph{Manifestations of Berry's phase
    in molecules and in condensed matter}, J. Phys.: Condens. Matter
  \textbf{12}, R107\textendash R143 (2000).

\bibitem{Baroni:2001} S. Baroni, S. de Gironcoli, A. Dal Corso, and
  P. Giannozzi, \emph{Phonons and related crystal properties from
    density-functional perturbation theory},
  Rev. Mod. Phys. \textbf{73}, 515\textendash 562 (2001). 

\bibitem{QE:2009}P. Giannozzi \emph{et al.}, \emph{\textsc{Quantum
      ESPRESSO}: a modular and open-source software project for
    quantum simulations of materials}, J. Phys.: Condens. Matter
  \textbf{21}, 395502 (2009); \url{http://www.quantum-espresso.org}.

\bibitem{LAMMPS:1995}CMD simulations have been performed using the
  LAMMPS code, see: S. Plimpton, \emph{Fast parallel algorithms for
  short-range molecular dynamics}, J. Comp. Phys. \textbf{117},
  1\textendash 19 (1995); \url{http://lammps.sandia.gov}.

\bibitem{PP}Norm-conserving PP's from the \textsc{Quantum ESPRESSO}
  public repository
  (\url{http://pseudopotentials.quantum-espresso.org}) were used. The
  PP datasets used for Ar, O, and H are \texttt{Ar.pz-rrkj.UPF},
  \texttt{O.pbe-hgh.UPF}, and \texttt{H.pbe-vbc.UPF}, respectively.
  PW's up to a kinetic-energy cutoff of 24 Ry for Ar and 80 Ry for
  water were included in the basis set.

\bibitem{Grossman2004} J.C. Grossman, E. Schwegler, E.W. Draeger,
  F. Gygi, and G. Galli, \emph{Towards an assessment of the accuracy
    of density functional theory for first principles simulations of
    water}, J. Chem. Phys. \textbf{120}, 300\textendash 311 (2004).

\bibitem{Silvestrelli:1997} P. Silvestrelli, M. Bernasconi, and
  M. Parrinello, \emph{Ab initio infrared spectrum of liquid water},
  Chem. Phys. Lett. \textbf{277}, 478\textendash 482 (1997). 

\bibitem{Marcolongo:2014} A. Marcolongo, \emph{Theory and ab initio
    simulation of atomic heat transport}, SISSA PhD thesis (2014),
  \nobreak{\url{http://cm.sissa.it/thesis.php/2014/marcolongo}}

\bibitem{Expt:H2O} N. Matsunaga and A. Nagashima, \emph{Transport
    properties of liquid and gaseous D$_2$O over a wide range of
    temperature and pressure}, J. Phys. Chem. Ref.  Data \textbf{12},
  933\textendash 966 (1983); M.L.V. Ramires\textbf{ }\emph{et
    al.}\textbf{ }, \emph{Standard reference data for thermal
    conductivity of water}, J. Phys. Chem. Ref. Data\textbf{ 24},
  1377\textendash 1381 (1994).

\end{thebibliography}
\end{document}